\documentclass[pra,twocolumn,amsmath,amssymb,aps,superscriptaddress]{revtex4-1}

\usepackage{graphicx}
\usepackage{dcolumn}
\usepackage{physics}
\usepackage{bm}

\begin{document}


\title{Quantum coherence in the Dynamical Casimir Effect}

\author{D. N. Samos-S\'aenz de Buruaga}
 \email{nadirbon@gmail.com}
\affiliation{%
Facultad de Ciencias F\'isicas, Universidad Complutense de Madrid, Plaza Ciencias, 1 Ciudad Universitaria 28040 Madrid (Spain)}%
\author{Carlos Sab\'in}
\email{csl@iff.csic.es}
\affiliation{
 Instituto de F\'isica Fundamental, CSIC, Serrano, 113-bis, 28006 Madrid (Spain)
}%

\date{\today}

\begin{abstract}
We propose to use quantum coherence as ultimate proof of quantum nature in the radiation that appears by means of the Dynamical Casimir Effect in experiments with superconducting microwave waveguides. We show that, unlike previously considered measurements such as entanglement and discord, quantum coherence does not require a threshold value of the external pump amplitude and is highly robust to thermal noise. 
\end{abstract}

\pacs{Valid PACS appear here}
\maketitle


\section{Introduction}

The dynamical Casimir effect (DCE) is a vacuum amplification process  that can produce pairs of photons from quantum vacuum fluctuations by means of the  modulation of a boundary condition -e. g. a mirror- at relativistic speeds. Theoretically discovered in 1970 \cite{moore}, the experimental demonstration was impossible until 2011 when Wilson et. al. \cite{wilson} managed to create in the laboratory a boundary condition equivalent to the one created by a mirror, as well as move it at relativistic speeds. This was achieved by means of high-frequency modulation of the external magnetic flux threading a superconducting quantum interferometric device (SQUID), which interrupted a superconducting coplanar microwave guide.
The effective motion induced by the magnetic field generated the expected microwave photon pairs out of an initial quasi-vacuum state. While some hints of the truly quantum nature of the radiation were provided in \cite{wilson}, the presence of an initial temperature leads naturally to the question of how unambiguously discriminate between the experimental radiation and a mere parametric amplification of thermal noise.

By a theoretical approach it has been shown that there exist measurements of quantum correlations which allow to quantify the non-classicality of the DCE radiation even when a background thermal noise is considered \cite{johan2,sabin,steering}. In all the cases, the proposed measurements are non-zero only above a critical value of the amplitude of the external pump -and thus the amplitude of the motion- which is needed to overcome the thermal noise. This feature challenges the experimental observation of quantum correlations in DCE radiation for realistic parameters. Therefore, the question of the quantum nature of DCE radiation in this setup remains open --- note however that quantum correlations have been detected in an analogue effect where the radiation is generated by the modification of the effective speed of light, not the boundary conditions \cite{sorin}.

Last years have witnessed an increasing interest in capturing forms of quantum correlations and other quantum features that are more general than quantum entanglement. In particular, quantum coherence is gaining a great deal of attention as a signature of quantum nature and as a resource for quantum technologies \cite{reviewcoherence,baumgratz}. 
In this work  we have studied a recently proposed measurement of quantum coherence for gaussian states in continuous-variables  \cite{jianwei}. We show that DCE radiation displays quantum coherence for any value of the pump amplitude in the typical experimental regime of temperatures. Unlike quantum correlations such as entanglement and discord, quantum coherence exhibits a remarkable robustness again thermal noise, making it a suitable figure of merit of the quantum character of DCE radiation.

The work is structured in the following way. In Section II we review the theoretical details of DCE in superconducting circuits experiments. In Section III we briefly introduce some aspects of the covariance matrix (CM) formalism for Gaussian states. We use the results of these sections to write the experimental DCE-radiation state in the CM formalism in Section 4. In section V we compute the quantum coherence of this state and discuss its behaviour for realistic experimental parameters, comparing it with quantum entanglement and discord. We conclude in Section VI with a summary of our results.    


\section{The Dynamical Casimir Effect in superconducting transmission lines}

The DCE predicts that photons can be created from vacuum fluctuations when the boundaries of the field are time-dependent. 
From a physical point of view, we can understand it by recalling that the quantum field vacuum is a sea of  virtual photons and energy fluctuations. Thus, we can think of the DCE as the incapacity of the vacuum state to accommodate to a fast changing boundary; as a consequence, some of those virtual photons are ``kicked'' to reality.

 In 2011 the first experimental observation of this phenomenon was reported \cite{wilson}. The difficulty of accelerating a mirror to relativistic speeds was circumvented by simulating a moving mirror with a superconducting quantum interferometric device (SQUID) interrupting a superconducting transmission line.
 
Let us review the details of the model. Quantizing the electromagnetic field by the flux operator $\Phi(x,t)$ confined in this transmission line and solving the Klein-Gordon equation in the plane-waves basis we get \cite{johan,johan2,johan1}:
 \begin{eqnarray}
 \label{flujo}
 \Phi(x,t)&=&\\ \nonumber
 &\sqrt{\frac{\hbar Z_0}{4\pi}}&\int \frac{d\omega}{\sqrt{\omega}}(\hat{a}(\omega)e^{i(-\omega t+kx)}+\hat{b}(\omega)e^{i(-\omega t-kx)}),
 \end{eqnarray}
where $a(\omega)$ is the annihilation operator of photons that fall on (input) the mirror and $b(\omega)$ is the annihilation operator of those who move away from the mirror (output), $k=\frac{\omega}{v}$ where $v$ is the speed of light in the line and $Z_0=\sqrt{L_0/C_0}$ is the characteristic impedance. Note that we are interpreting the creation operators as annihilation operators of negative frequency. As shown in \cite{johan}, for sufficiently large SQUID plasma frequency, such that the charging energy is small compared with the Josephson energy $E_J(t)=E_J[\Phi(x,t)]$, the SQUID provides a boundary condition to the flux field  analogous to the one produced by a mirror:
\begin{equation}
\label{boundary}
\frac{(2\pi)^2}{\phi^2_0}\Phi(0,t)+\frac{1}{L_0}\left.\frac{\partial\Phi(x,t)}{\partial x}\right|_{x=0}=0,
\end{equation}
where $\Phi_0=\frac{h}{2e}$ is the magnetic flux quantum, $L_0$ is the characteristic inductance per unit length and we have neglected an additional term associated with the capacitance.
Inserting Eq.(\ref{flujo}) on the boundary condition Eq.(\ref{boundary}) we have:
\begin{eqnarray}
\nonumber
\int_0^\omega \frac{d\omega}{\sqrt{\omega}}ik\frac{L_{J}(t)}{L_0}(\hat{b}(\omega)&-&\hat{a}(\omega))e^{-i\omega t}=\\
=\int_0^\omega \frac{d\omega}{\sqrt{\omega}}(\hat{a}(\omega)&+&\hat{b}(\omega))e^{-i\omega t},
\end{eqnarray}
where $L_J(t)=\frac{\phi^2_0/(2\pi)^2}{E_J(t)}$ is the tunable Josephson inductance.
Inserting the Josephson energy $E_J(t)$ in the latter,  we obtain a Bogoliubov transformation between the input and output modes. If we consider a constant Josephson energy $E_J^0$ --which corresponds to a constant flux $E_J^0=E_J[\Phi^0(x)]$-- and integrate over time, we get the transformation:
\begin{equation}
\label{bogoconst}
\hat{b}(\omega)=-\frac{1+ik\frac{L_J^0}{L_0}}{1-ik\frac{L_J^0}{L_0}}\hat{a}(\omega)\approx -\exp\{2ik\frac{L_J^0}{L_0}\}\hat{a}(\omega)
\end{equation} 
where we have assumed that $k\frac{L_J^0}{L_0}<<1$. Given that the reflection coefficient of a short-circuited coplanar microwave guide of length $L$ is $-e^{2ikL}$, we can define an effective length $L_{eff}^0=\frac{L_J^0}{L_0}$ and conclude that:
\begin{equation}
b(\omega)=-\exp\{2ikL_{eff}^0\}\hat{a}(\omega)=R(\omega)\hat{a}(\omega).
\end{equation}
Considering now a time-dependent effective Josephson energy $E_J(t)=E_J^0+\delta E_J(t)$ it can be shown \cite{johan} that we can write:
\begin{eqnarray}
\label{generalbog}
\hat{b}(\omega)&=&R(\omega)\hat{a}(\omega)-\int_{-\infty}^{\infty}d\omega' S(\omega,\omega')\times \\ \nonumber
&\times & \Theta(\omega)\left(\hat{a}(\omega)+\hat{b}(\omega)\right)+\Theta(-\omega)\left(\hat{a}(\omega)+\hat{b}(\omega)\right)^\dagger,
\end{eqnarray}
where $\Theta(\omega)$ is the Heaviside step function and 
\begin{equation}
S(\omega,\omega')=\frac{\sqrt{|\omega/\omega'|}}{2\pi(1-ikL_{eff}(0))}\int_{-\infty}^{\infty}dt e^{-i(\omega-\omega')t}\frac{\delta E_J(t)}{E_J^0}.
\end{equation}

In particular, given a sinusoidal modulation with \emph{driven frequency} $\omega_d/2\pi$ and normalized amplitude:
\begin{equation}
\label{joseph} E_J(t)=E_J^0[1+\epsilon\sin{\omega_dt}],\quad \epsilon<<1,
\end{equation}
we get, after integration of Eq.(\ref{generalbog}):
\begin{eqnarray}
\label{bogososcil}
\nonumber
\hat{b}(\omega)&=&R(\omega)\hat{a}(\omega)+i\frac{\delta L_{eff}}{v}\left(\sqrt{\omega(\omega+\omega_d)}\hat{a}(\omega+\omega_d)\right.\\ \nonumber
 &+&\sqrt{|\omega(\omega-\omega_d)|}\times\left[\Theta(\omega-\omega_d)\hat{a}(\omega-\omega_d)+\right.\\
 &\quad &\quad\quad\quad\quad\quad\quad\quad +\left.\left.\Theta(\omega_d-\omega)\hat{a}^\dagger(\omega_d-\omega)\right]\right),
\end{eqnarray}
where $\delta L_{eff}=\epsilon L_{eff}^0$ is the effective length modulation. 

Considering 
the effective velocity: $$v_{eff}=\epsilon L_{eff}^0\omega_d=\delta L_{eff}\omega_d,$$ we can say that if $v_{eff}$ is a significant fraction of $v$, radiation with a squeezing spectrum \cite{johan} will appear.

Since the photon-flux density spectrum has a parabolic shape with a maximum in $\frac{\omega_d}{2}$ and the photons are created in pairs, the output field is correlated at modes with angular frequencies $\omega_+$ and $\omega_-$ such that $\omega_++\omega_-=\omega_d$. We will consider that:
\begin{equation}
\omega_\pm = \frac{\omega_d}{2}\pm\delta\omega,
\end{equation}
where $\delta\omega$ is a small detuning that allows to write Eq.(\ref{bogososcil}) as:
\begin{equation}
b_{\pm}=-a_\pm-i\frac{\delta L_{eff}}{v}\sqrt{\omega_+\omega_-}a^\dagger_{\mp},
\end{equation}
where $b_{\pm}=b(\omega_\pm)$ and $a_{\pm}=a(\omega_\pm)$.  If the detuning is sufficiently small, we can write: $$\omega_-\approx\omega_+\approx\frac{\omega_d}{2},$$ and, consequently, define a small parameter $f$:
\begin{equation}
\label{parameter}
\frac{\delta L_{eff}\sqrt{\omega_+\omega_-}}{v}\approx\frac{\delta L_{eff}\omega_d}{2v}=\frac{v_{eff}}{2v}=f.
\end{equation}
So, finally, the relation between the output and input fields is the following:
\begin{equation}
\label{bogos}
b_{\pm}=-a_\pm-ifa^\dagger_{\mp}.
\end{equation}

In the next section we will introduce the formalism of the covariance matrix for the analysis of Gaussian states.
  
\section{Gaussian States. Covariance Matrix}
 
In this section we will review the covariance matrix formalism, which is suitable for the analysis of Gaussian states. 
Gaussian states appear in any quantum system which can be described by a quadratic bosonic Hamiltonian. Therefore, they play a central role in quantum optics and they are the core of quantum information with continuous variables \cite{adesso}. 
 
In particular, the electromagnetic field is given by a quadratic bosonic Hamiltonian which describes a system of N harmonic oscillators:
\begin{equation}
\label{hamilt}
\hat{H}=\sum_{k=1}^N\hbar\omega_k\left(\hat{a}^\dagger_k\hat{a}_k+\frac{1}{2}\right).
\end{equation}

The Hamiltonian (\ref{hamilt}) describes the prototype of a continuous-variable system which has an infinite-dimensional separable Hilbert space $\mathcal{H}= \bigotimes_{k=1}^N\mathcal{H}_k $ where $\mathcal{H}_k$ is the infinite-dimensional Fock space spanned by the annihilation/creation operators of the $k$ mode. 
The creation/annihilation operators can be arranged in a vectorial operator $\hat{\mathbf{b}}:=(\hat{a}_1,\hat{a}^\dagger_1,...,\hat{a}_N,\hat{a}^\dagger_N)^T$ which must satisfy the bosonic commutation relations:
\begin{equation}
[\hat{b}_i,\hat{b}_j]=\Omega_{ij}\quad(i,j=1,...2N).
\end{equation}
In the same way that a global metric allows to define a metric vector space, a linear dynamical system has a symplectic form $\Omega$ that does not depend on the coordinates, allowing us to study it by a symplectic vector space called phase space $\Gamma=(\mathbb{R}^{2n},\Omega)$. The symplectic form has a direct sum structure: 
\begin{equation}
\mathbf{\Omega}:=\bigoplus_{k=1}^N\omega=
 \begin{pmatrix}
  \omega & &  & \\
   &   & \ddots & \\
   &  &  & \omega 
 \end{pmatrix}\quad \omega:=\begin{pmatrix}
 0&1\\
 -1&0\end{pmatrix}
\end{equation}
that transfers the same structure to the phase space $\Gamma=\bigoplus_k\Gamma_k$  with $\Gamma_k=\Gamma=(\mathbb{R}^{2n},\omega)$.

Making a Cartesian decomposition of the bosonic field operators we get the quadratures of the field:
\begin{equation}
\label{quadratures}
\hat{q}_k:=\frac{1}{\sqrt{2}}\left(\hat{a}_k+\hat{a}^\dagger_k\right)\quad\hat{p}_k:=\frac{i}{\sqrt{2}}\left(\hat{a}^\dagger_k-\hat{a}_k\right),
\end{equation}
that represent dimensionless canonical observables of the system. If we arrange them in a vector,
\begin{equation}
\hat{\mathbf{x}}:=(\hat{q}_1,\hat{p}^\dagger_1,...,\hat{q}_N,\hat{p}^\dagger_N)^T.
\end{equation}
Now we can define 
N-mode Weyl operators \cite{adesso}. Given $\boldsymbol{\xi}\in\Gamma$,
\begin{equation}
\hat{\mathbf{D}}_{\xi}:=e^{i\hat{\mathbf{x}}^T\mathbf{\Omega}\boldsymbol{\xi}}.
\end{equation}
It is possible to give a complete description of the quantum infinite dimensional states of a system in the framework of the phase space by one of its  s-ordered \emph{characteristic function}:
\begin{equation} \chi_s(\boldsymbol{\xi})=\textit{Tr}\left[\rho\hat{\mathbf{D}}_{\xi}\right]e^{\frac{s||\boldsymbol{\xi}||}{2}},
\end{equation}
 where $||\cdot||$ is the Euclidean norm. Moreover, the complex Fourier transforms of these characteristic functions are another valid description of the quantum states and are called \emph{quasi-probability distributions}. The constant $s$ is deeply related with the ordering of the creation/annihilation operators. If we choose $s=0$ we have the Wigner function $W$ where the operators are symmetrically ordered:
\begin{equation}
W(\boldsymbol{\xi})=\frac{1}{\pi^2}\int_{\mathbb{R}^{2N}}d^{2N}\mathbf{x}\chi_0(\boldsymbol{\xi})e^{i\mathbf{x}^T\Omega\boldsymbol{\xi}}.
\end{equation}
A quantum state is said to be Gaussian if its characteristic function is Gaussian. Thus, a Gaussian state can be written in terms of two objects:
\begin{itemize}
\item The first moment vector $\langle \hat{\mathbf{x}}\rangle$, which can be adjusted by applications of the single-mode Weyl operator, so it can always be set to $0$ without any loss of generality -as we will use later on.
\item The second moment matrix or covariance matrix (CM), which is the most important object in this section:
\begin{equation}
\mathbf{\sigma}_{ij}=\frac{1}{2}\langle \hat{x}_i\hat{x}_j+ \hat{x}_j\hat{x}_i\rangle-\langle \hat{x}_i\rangle\langle\hat{x}_j\rangle.
\end{equation}
All the information of the Gaussian states is encoded in this $2N\times2N$ matrix, although it is necessary to impose some constraints to this matrix, in order to the density matrix be positive-semi definite (\emph{bona fide}) $\mathbf{\sigma}+i\Omega\geq0$. 
\end{itemize}
It is convenient to write the CM in blocks: each of the diagonal $\mathbf{\sigma}_k$ is respectively the local CM corresponding to the reduced state of the $k$ mode, and the off-diagonal matrices $\mathbf{\epsilon}_{i,j}$ encode all the correlations (classical and quantum) between the subsystems $i$ and $j$:
\begin{equation}
\label{covariance}
\boldsymbol{\sigma}=
\begin{pmatrix}
\boldsymbol{\sigma}_1 & \boldsymbol{\epsilon}_{1,2j} & \ldots & \boldsymbol{\epsilon}_{1,N}\\
\boldsymbol{\epsilon}^T_{1,2j} &\ddots &\ddots & \vdots\\
\vdots & \ddots &\ddots & \boldsymbol{\epsilon}_{N-1,N}\\
\boldsymbol{\epsilon}^T_{1,N}& \ldots & \boldsymbol{\epsilon}^T_{N-1,N}& \boldsymbol{\sigma}_N
\end{pmatrix}.
\end{equation}

As we said before, because of the symplectic form furnishes with a rich geometry the manifold, it is possible to associate unitary operations with symplectic operations, that preserve the symplectic form. One of the most important symplectic transformation $\mathbf{S}$ is the one that diagonalize the CM,
\begin{equation}
\boldsymbol{\sigma}=\mathbf{S}^T\boldsymbol{\nu}\mathbf{S}\quad,\boldsymbol{\nu}=\bigoplus_{k=1}^N\begin{pmatrix}
\nu_k&0\\
0&\nu_k
\end{pmatrix},
\end{equation}
where $\nu_k$ are called \emph{symplectic eigenvalues}, which can be computed as the absolute value of the orthogonal eigenvalues of the matrix $i\Omega\boldsymbol{\sigma}$. Furthermore, the Williamson theorem ensures that there exist global symplectic invariants such as the determinant of the CM $Det\boldsymbol{\sigma}=\prod_{k=1}^N\nu_k^2$ and the \emph{Seralian} $\Delta(\boldsymbol{\sigma})=\sum_{k=1}\nu^2_k$ i.e. the sum of all the determinants of the $2\times2$ sub-matrices. As we will see, these invariants are crucial in many computations because it is possible to write physical magnitudes in terms of them. 

For instance, it can be shown \cite{serafini} that the Von Neumann entropy takes the form:
\begin{eqnarray}
\label{entropy}
&S_V(\rho)&= \sum_{k=1}^N h(\nu_k)=\\ \nonumber
&\sum_{k=1}^N&\left(\frac{\nu_k+1}{2}\right)\log\left(\frac{\nu_k+1}{2}\right)-\left(\frac{\nu_k-1}{2}\right)\log\left(\frac{\nu_k-1}{2}\right),
\end{eqnarray}
where 
\begin{equation}
\label{eq}
h(x)=\frac{x+1}{2}\log{\frac{x+1}{2}}-\frac{x-1}{2}\log{\frac{x-1}{2}}.
\end{equation}
We will use this magnitude later on.

In the next section, we will describe the DCE in the CM formalism, by applying the Bogoliubov transformation obtained in Section II to a realistic initial quasi-vacuum state.
\section{DCE in the CM formalism: initial and final states}
In this section,we make use of the results of the previous sections in order to describe the DCE in the CM formalism.

We start by considering a proper initial state. First of all, we shall restrict ourselves to a pair of central modes $-,+$ with close frequencies -- as discussed in Section II.Then, although the ideal observation of the DCE would occur in a vacuum state, unfortunately it is impossible to get it experimentally, so we assume a quasivacuum state characterized by a small fraction of thermal photons $n_+^{th},n_-^{th}$. Because it is a coherent Gaussian state, the two-mode CM associated $\boldsymbol{\sigma}_i$ is proportional to the identity matrix \cite{adesso}:
\begin{equation}
\boldsymbol{\sigma}_i:=\frac{1}{2} \left(
\begin{array}{cccc}
 2 n_-^{th}+1 & 0 & 0 & 0 \\
 0 & 2 n_-^{th}+1 & 0 & 0 \\
 0 & 0 & 2 n_+^{th}+1 & 0 \\
 0 & 0 & 0 & 2 n_+^{th}+1 \\
\end{array}
\right).
\end{equation}

Using Eq.(\ref{quadratures}) and Eq. (\ref{bogos}) we can obtain the output quadratures in terms of the input ones $q_i$:
\begin{eqnarray}
\label{quadraout}
q_{\pm}&=&-(q_{i\pm}+fp_{i\mp})\\
p_{\pm}&=&-(p_{i\pm}+fq_{i\mp})
\end{eqnarray}

 With $\boldsymbol{\sigma}_i$ and the quadratures Eq.(\ref{quadraout}) we get the output CM $\boldsymbol{\sigma}$:
 \begin{equation}\label{eq:out1}
 \boldsymbol{\sigma}=\begin{pmatrix}
 \boldsymbol{\sigma}_+&\boldsymbol{\epsilon}_{-+}\\
 \boldsymbol{\epsilon}_{+-}&\boldsymbol{\sigma}_-
 \end{pmatrix}.
 \end{equation}
 Where:
 \begingroup\makeatletter\def\f@size{8}\check@mathfonts
 \begin{eqnarray}\label{eq:out2}
 \nonumber
 &\boldsymbol{\sigma}_- &=\frac{1}{2}\begin{pmatrix}
  (2 n_+^{th}+1) f^2+2 n_-^{th}+1 & 0\\
   0 & (2 n_+^{th}+1) f^2+2 n_-^{th}+1
  \end{pmatrix}\\ \nonumber
 &\boldsymbol{\sigma}_+&=\frac{1}{2}\begin{pmatrix}
  (2 n_-^{th}+1) f^2+2 n_+^{th}+1 & 0\\
  0 & (2 n_-^{th}+1) f^2+2 n_+^{th}+1 
  \end{pmatrix}\\\nonumber
 &\boldsymbol{\epsilon}_{-+}&=\frac{1}{2}\begin{pmatrix}
  0  & 2 f (n_+^{th}+n_-^{th}+1)\\ 
  2 f (n_+^{th}+n_-^{th}+1) & 0 
  \end{pmatrix}\\ 
  &\boldsymbol{\epsilon}_{+-}&=\boldsymbol{\epsilon}^T_{-+}.  
 \end{eqnarray}\endgroup
Finally, as we are considering small detuning $\omega_+\approx\omega_-\approx\frac{\omega_d}{2}$ we have:
\begin{equation}
n_-^{th}\approx n_+^{th}\approx n^{th}.
\end{equation} 

Eqs. (\ref{eq:out1}) and (\ref{eq:out2}) represent the state of the DCE  radiation. The main task is thus how to find a proper indicator of its quantum nature, as opposed, for instance to mere thermal radiation. We will adress this issue in the next section.

\section{Quantum correlations and quantum coherence in the DCE}

We have seen that the DCE is an effect that creates radiation with a genuinely quantum nature, due to its origin in the virtual particles of the quantum vacuum. Because of that, this radiation must possess quantum correlations (QC).
Quantum systems in a composite state display QC if they are not classically correlated. A composite system will present classical correlations if at least one of its constituents is a classical system whose states can be represented using a orthonormal basis and it has a tensor product structure. For example $\rho_{AB}=\sum_ip_i\ket{i}_A\bra{i}_A\otimes\rho_B^{(i)}$  is classically correlated with respect to subsystem A (See \cite{adesso1}).
 
There are several kinds of QCs, as we will see in the following.
 
\subsection{Quantum entanglement}

Entanglement is the QC's tip of the iceberg, and it is a direct consequence of the superposition principle and the tensor structure of the Hilbert space.
One state of a bipartite system will present entanglement if it cannot be prepared by local operations, i.e. it is not separable:
\begin{equation}
\boldsymbol{\rho_{AB}}\neq\sum_i p_i\boldsymbol{\rho}_A^{(i)}\otimes\boldsymbol{\rho}_B^{(i)},
\end{equation}
with $\boldsymbol{\rho}_A^{(i)}$, $\boldsymbol{\rho}_B^{(i)}$ quantum states of subsystem A and B respectively.

In the phase space framework, a Gaussian state of a bipartite system with CM $\boldsymbol{\sigma}$ is separable if the partially transposed CM $\tilde{\boldsymbol{\sigma}}$ verifies $\tilde{\boldsymbol{\sigma}}+i\Omega\geq0$. The partial transpose operation in the phase space amounts to a mirror reflection of the mode's momentum operators of one subsystem. The logarithmic negativity quantifies the violation of this criterion, and in the case of only two modes is defined as:
\begin{eqnarray}
\label{neglog}
E(\boldsymbol{\sigma}) = \operatorname{Max}[0,-\log{2\tilde{\nu}_-}],
\end{eqnarray}
where $\tilde{\nu}_-$ is the smallest symplectic eigenvalue of $\tilde{\boldsymbol{\sigma}}$.

We now compute this magnitude for the DCE radiation state given by Eqs. (\ref{eq:out1}) and (\ref{eq:out2}). Taking a Taylor series expansion both for $f$ and $n^{th}$, we obtain
\begin{equation}
\tilde{\nu}_-=\frac{1}{2} (-1 + f)^2 (1 + 2 n^{th}).\label{eq:smallest}
\end{equation}
Considering Eqs.(\ref{neglog}) and (\ref{eq:smallest}) we obtain:
\begin{equation}
\label{nega}
E(f,n^{th})=\max\left[0,2(f-n^{th})\right].
\end{equation}

\subsection{Quantum discord}

Discord is a more general form of QC, which is displayed by some mixed states that are separable -therefore, non-entangled- but whose correlations are still non-classical. It can be quantified by the difference between total and classical correlations, which can be characterized by two different quantum versions of two classically equivalent definitions of the mutual information. 

In the CM formalism of Gaussian states, it is shown in Refs.\cite{giorda, adesso2,pirando} that the quantum discord of a bipartite $\boldsymbol{\sigma}$  has the following form:
\begin{eqnarray}\label{eq:discgauss}
D(\boldsymbol{\sigma})&=&h(\sqrt{\det{\boldsymbol{\sigma}_{2}}})-h(\nu_1)-h(\nu_2)\\ \nonumber
&+&
h(\frac{\sqrt{\det{\boldsymbol{\sigma}_1}}+2\sqrt{\det{\boldsymbol{\sigma}_1}\det{\boldsymbol{\sigma}_2}}+2\det{\boldsymbol{\epsilon}_{12}}}{1+2\sqrt{\det{\boldsymbol{\sigma}_2}}}),
\end{eqnarray}
where $h(x)$ is given in (\ref{eq}) and $\nu_i$, $i=1, 2$ are the symplectic eigenvalues.

A perturbative expression for the quantum discord of the DCE radiation can be obtained using Eqs. (\ref{eq:out1}) and (\ref{eq:out2}) and computing its symplectic eigenvalues:
\begin{eqnarray}\label{eq:symplecs}
\nu_- =\nu_+=\frac{1}{2}(1-f^2) (1 + 2 n^{th}).
\end{eqnarray}
Please note the difference with Eq. (\ref{eq:smallest}).\\
Inserting everything in Eq. (\ref{eq:discgauss}) and following the perturbative approach we obtain \cite{sabin}:
\begin{equation}
\label{discod}
D(f,n^{th})=\max \left[0,f^2-\frac{(n^{th})^2}{2}\right].
\end{equation}

Taking into account the expressions obtained for negativity and discord, it is obvious that -- for a given $n^{th}$ -- there exists a threshold depending on the small parameter $f$ which discriminates whether these QC overcome the thermal noise or not. These thresholds were obtained and compared in \cite{sabin}. In particular, it was shown that discord beats the thermal noise more easily than entanglement. In the following, we will show that the behaviour of the quantum coherence is dramatically different.
Finally, note that we are only computing the discord up to the leading order in perturbation theory. Higher-order terms might show up in a non-pertubative computation -- which would require a non-perturbative covariance matrix to begin with--, making discord non-zero for any non-zero value of the squeezing \cite{giordaallegroparis, blandino}. Please note that other \textit{bona fide} measurements of quantum discord might well display a different perturbative behaviuor.

\subsection{Quantum coherence}

Quantum coherence amounts to superposition with respect to a fixed orthonormal basis: one state $\rho$ is incoherent in a fixed basis if it is diagonal. Therefore, there are mixed states which do not possess entanglement or discord but display some form of quantum coherence. The role of quantum coherence in modern quantum technologies such as quantum thermodynamics, quantum metrology or quantum biology is currently the subject of intense research -- see the recent review \cite{reviewcoherence}. 

A coherence measure $C(\rho)$, where $\rho$ is now a $n$-mode Gaussian state, has been recently given in terms of the covariance matrix and displacement vectors in \cite{jianwei}: $C(\rho)=\textit{inf}\{S(\rho||\delta)\}$ where $\delta=\bigotimes_{k=1}^n\delta^k_{th}(\bar{n}_k)$ is a tensor product of reduced thermal states with respect each mode $k$.
Considering the Von-Neumann entropy of the system in terms of the symplectic eigenvalues (\ref{entropy}) and the mean occupation value:
\begin{equation}
\bar{n}_k=\frac{1}{2}\left(\boldsymbol{\sigma}^{11}_{k}+\boldsymbol{\sigma}^{22}_{k}+[d_1^{(k)}]^2+[d_2^{(k)}]^2-1\right),
\end{equation}
-- where $\boldsymbol{\sigma}^{ij}_{k}$  is the $ij$ element of the $k$ mode CM and $[d_i^{(k)}]$ is the $i$ first statistical moment of the $k$ mode, which we will set to zero -- and performing an extremal calculus, it is possible to provide an analytical expression of the quantum coherence \cite{jianwei} :
\begin{eqnarray}
\label{coherence}
C(\boldsymbol{\sigma})&=&\\ \nonumber
&-&S(\boldsymbol{\sigma})+\sum_{k=1}^n(\bar{n}_k+1)\log_2(\bar{n}_k+1)-\bar{n}_k\log_2\bar{n}_k.
\end{eqnarray}
Thus, considering again the DCE state in Eqs. (\ref{eq:out1}) and (\ref{eq:out2}), the symplectic eigenvalues from Eq. (\ref{eq:symplecs}) and:
\begin{eqnarray}\label{eq:number}
\bar{n}_-&=&\bar{n}_+=n^{th} + f^2 (1/2 + n^{th}),
\end{eqnarray}
and inserting everything in Eq. (\ref{coherence}) -- always within the perturbative approach--, we find that the quantum coherence in the experimental basis is: 
\begin{eqnarray}
\label{cohe}
C(f,n^{th})&=&\\ \nonumber
&\max& \left[0,-2 f^2 (-n^{th}+(2 n^{th}+1)\log (n^{th}))\right],
\end{eqnarray}
which is the main result of this paper.

\begin{figure}[h!]
\includegraphics[scale=0.5]{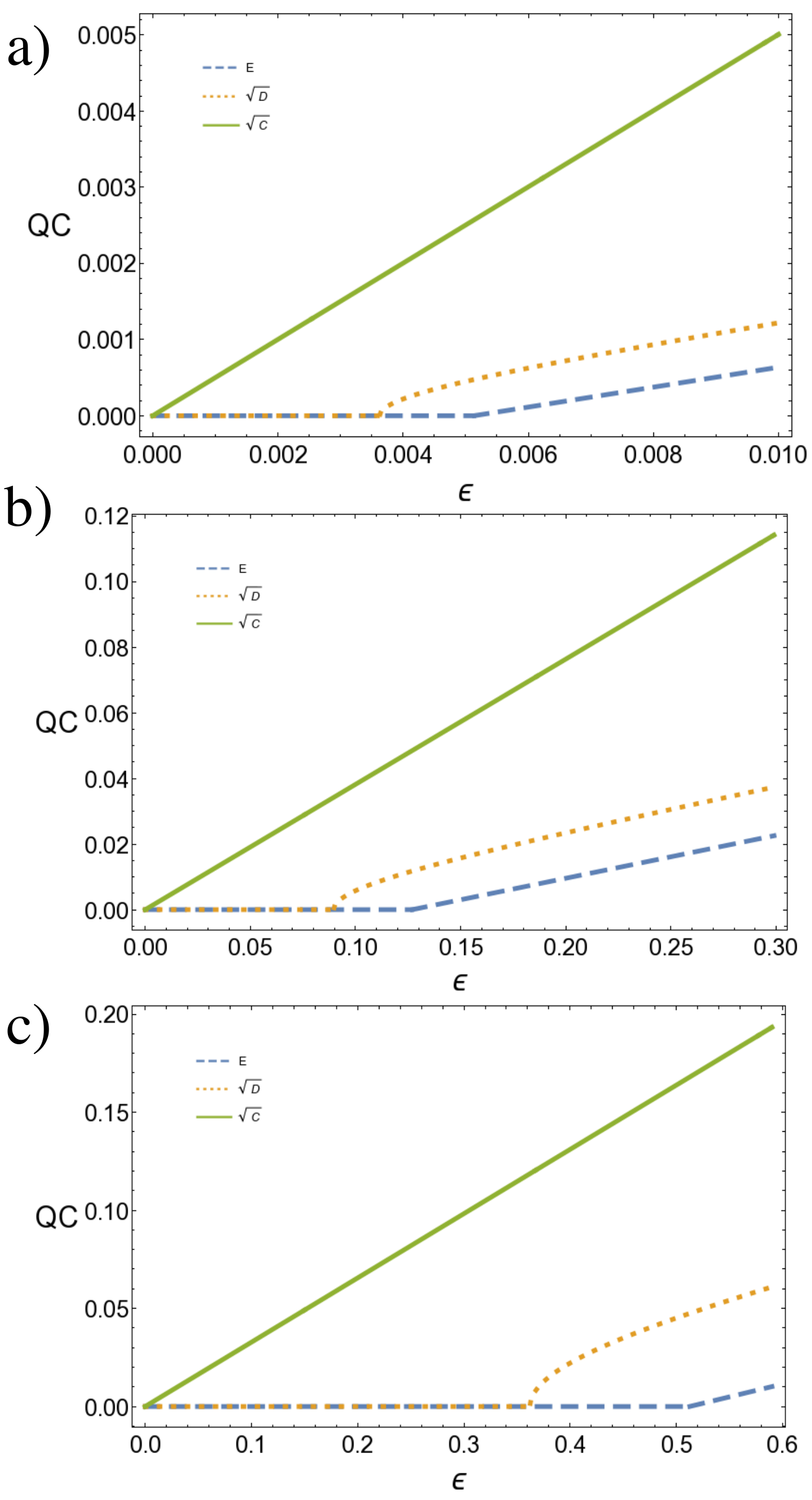}
\caption{Negativity ($E$), quantum discord $\sqrt{D}$ and coherence $\sqrt{C}$ as a function of the driving frequency for a given average input thermal photons a) corresponds to $T=30$mK ($n^{th}=6.71\times10^{-4}$), b) to $T=50$mK ($n^{th}=1.67\times10^{-2}$) and c) $T=70$mK ($n^{th}=6.71\times10^{-2}$). The plot is realized with experimental parameters $\omega_d=20\pi$ GHz, $L_{eff}^0=0.5$ mm and $v=1.2\times10^8$ m/s and the range of epsilon guarantees that the small parameter $f<0.08$ is well within the perturbative regime. While entanglement and discord have a threshold value of $\epsilon$ below which they are 0, quantum coherence is finite for any value of $\epsilon$. Note that $\epsilon=\frac{2\,v}{L^0_{eff}\omega_d}f$.}\label{fig:mak} 
\end{figure}

Therefore, the behaviour of the quantum coherence is different from the one of the quantum correlations considered: just by inspection of Eq.(\ref{cohe}) it is clear that there will be a positive amount of coherence for every $f\neq0$, i.e. for every driving amplitude. This is in agreement with the observation in \cite{jianwei} that any state with non-diagonal CM -- see Eq. (\ref{eq:out1}) -- possesses quantum coherence.  

\begin{figure}[t!]
\includegraphics[scale=0.5]{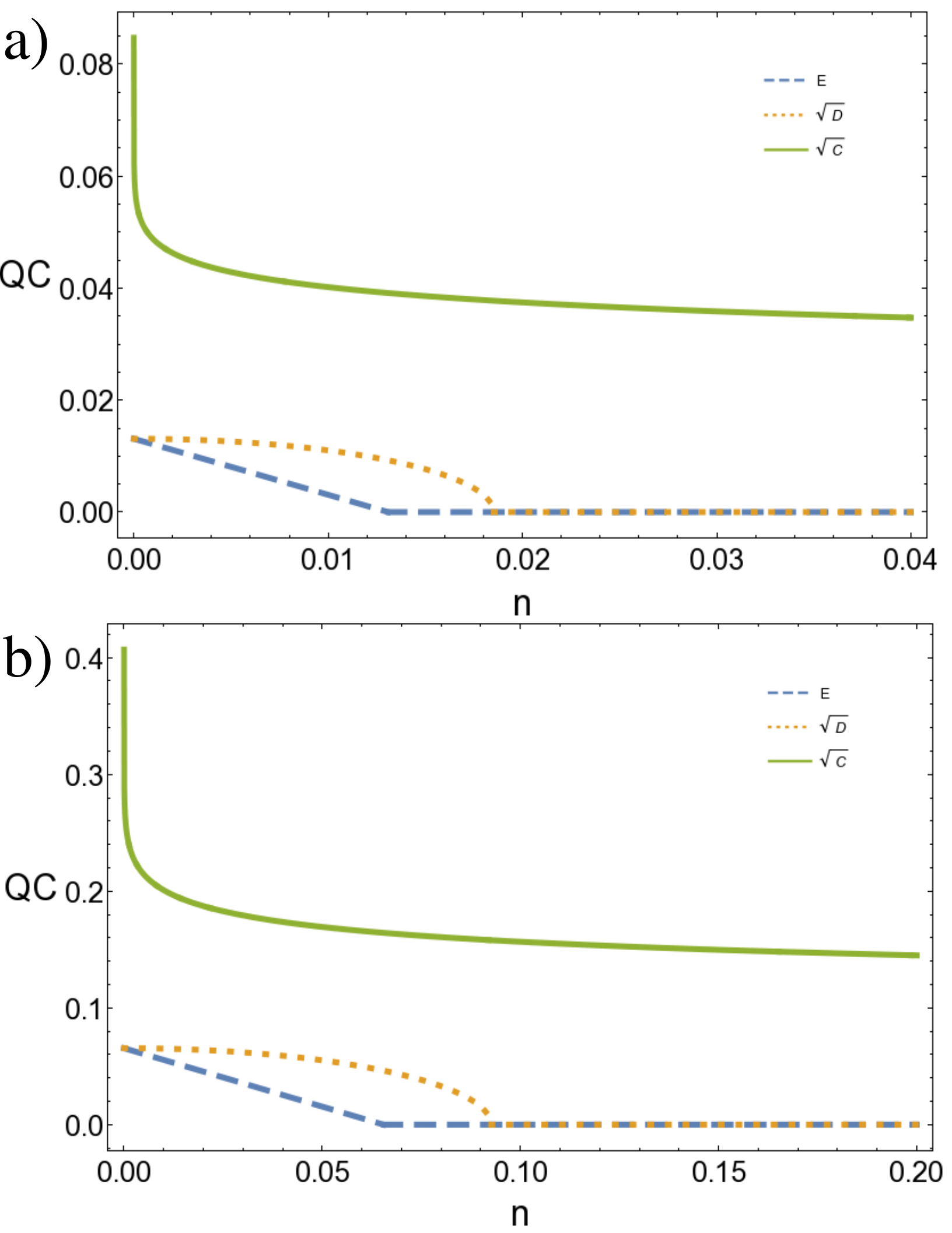}
\caption{Negativity ($E$), quantum discord $\sqrt{D}$ and coherence $\sqrt{C}$ as a function of the average input thermal photons for a given driving amplitude a) corresponds to $\epsilon=0.1$ and b) to $\epsilon=0.5$). The plot is realized with experimental parameters $\omega_d=20\pi$ GHz, $L_{eff}^0=0.5$ mm and $v=1.2\times10^8$ m/s. While quantum entanglement and quantum discord vanishes for a given thermal noise threshold, coherence does not disappear in that regime of temperatures and driving amplitudes. Also, quantum entanglement is less robust than discord: In top figure, entanglement vanishes at $T\approx48$ mK and discord survives up to $T\approx67$ mK, while in the bottom figure, the threshold is displaced to $T\approx70$ mK for the entanglement and $T\approx100$ mK for the discord. Note that coherence is always larger than the other two.}\label{fig:fig2}
\end{figure}

In order to further illustrate this, we have represented in figures Fig. \ref{fig:mak} and Fig. \ref{fig:fig2} the logarithmic negativity and the square roots of the discord and the coherence --the square roots are considered in order to ensure that we include the same perturbative orders in all the cases \cite{sabin,max}. The plots have been made considering realistic experimental values \cite{wilson} and without violating the range of validity of the perturbative and small-detuning approximations.

In Fig. \ref{fig:mak} we confirm that coherence is significantly different from 0 for any value of the external pump amplitude, a property which becomes very interesting for experimental temperatures of $50 \, \operatorname{mK}$ -- as in \cite{wilson} -- and above, since entanglement and discord require very high values of $\epsilon$ in order to be clearly non-zero.

In Fig. \ref{fig:fig2} it is shown how, for a fixed $\epsilon$, entanglement vanishes at lower temperatures than discord, while coherence does not vanish at all. Moreover, we find that quantum coherence is always greater than discord and  entanglement. 

All these results strongly suggest that quantum coherence could be used as a smoking gun for the quantum nature in the DCE radiation, since the experimental requirements for obtaining a DCE state with finite coherence are much less demanding than in the case of entanglement or discord.

\section{Conclusion} 

In this work we have studied theoretically the quantum coherence of the state resulting from performing a DCE experiment with a superconducting waveguide terminated by a SQUID. We have found that quantum coherence has a different behaviour of previously considered measurements of quantum correlations such as discord and entanglement, because it does not require a driving amplitude threshold to appear. Therefore we hope that these results could be a useful tool for the confirmation of the quantum nature of DCE radiation. A natural extension of this work would be the analysis of how the DCE coherence generated in superconducting circuits experiments can be exploited as a resource for quantum technologies such as quantum thermodynamics, quantum metrology or quantum biology, where coherence seems to play a pivotal role \cite{reviewcoherence}.

\section*{Acknowledgements}

Financial support from Fundaci\'on General CSIC (Programa ComFuturo) is acknowledged by C.S. Additional support from MINECO Project FIS2015-70856-P (cofinanced by FEDER funds)
and CAM PRICYT Project QUITEMAD+ S2013/ICE-2801 is also acknowledged.

\end{document}